# Data Acquisition and Database Management System for Samsung Superconductor Test Facility

Y. Chu, S. Baek, H. Yonekawa, A. Chertovskikh, M. Kim, J. S. Kim, K. Park,
S. Baang, Y. Chang, J. H. Kim, S. Lee, B. Lim, W. Chung, H. Park, K. Kim
Samsung Advanced Institute of Technology, Taejon, Korea

## Abstract

In order to fulfill the test requirement of KSTAR (Korea Superconducting Tokamak Advanced Research) superconducting magnet system, a large scale superconducting magnet and conductor test facility, SSTF (Samsung Superconductor Test Facility), has been constructed at Samsung Advanced Institute of Technology. The computer system for SSTF DAC (Data Acquisition and Control) is based on UNIX system and VxWorks is used for the real-time OS of the VME system. EPICS (Experimental Physics and Industrial Control System) is used for the communication between IOC server and client. A database program has been developed for the efficient management of measured data and a Linux workstation with PENTIUM-4 CPU is used for the database server. In this paper, the current status of SSTF DAC system, the database management system and recent test results are presented.

## 1 INTRODUCTION

The KSTAR device is a tokamak with a fully superconducting magnet system, which enables an advanced quasi-steady-state operation. The major radius of the tokamak is 1.8 m and the minor radius is 0.5 m with the elongation of 2. The superconducting magnet system consists of 16 TF (Toroidal Field) coils and 14 PF (Poloidal Field) coils. The arrangement of the KSTAR coil system is shown in Fig.1.

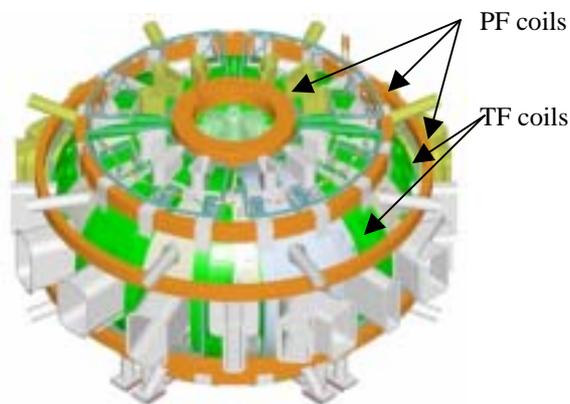

Fig.1 Arrangement of the KSTAR Magnet System

SSTF is constructed in order to test superconducting strands, conductors, and magnets. It consists of a vacuum cryostat system, a cryogenic cooling system, a background magnetic field generation system, a power supply system and a data acquisition and control system. Large Vacuum Cryostat (LVC) with diameter 6 m and height 7.9 m is located in a pit of 9 m x 9 m x 6 m. It is connected to three cold boxes, Helium path control valve box (CB#1), Helium flow rate control box (CB#2), and current lead box (CB#3), respectively. The cryogenic cooling system has two 200-watt Helium refrigerators and one 1000-watt Helium refrigerator, which is under construction. The power supply system also has a superconducting transformer type power supply, which is able to provide 50 kA to a superconducting CICC (Cable-In-Conduit Conductor) sample. The operation scenario of KSTAR device requires a steady state operation of TF coils and a fast ramping of PF coils. Thus, the test of KSTAR superconducting magnet and CICC also need to be performed under the condition of the fast varying magnetic field. The background magnetic field generation system is designed to provide a changing magnetic field of a 3 Tesla/second for 5 seconds and a 20 Tesla/second for 0.05 seconds. Therefore the data acquisition system has two modes, a high-speed data logging and a low-speed monitoring. The low-speed monitoring system is used to control slow-varying parameters such as Helium flow rate, temperature, and pressure, etc. However, a quench protection system, which is linked with the power supply system, and a magnetic measurement system require a high-speed data acquisition.

The computer system for SSTF DAC is based on UNIX and Linux system and VxWorks and RTEMS are used for the real-time OS of the VME system. The EPICS is the basic communication software for the monitoring of slow-varying parameters and RT-Linux system with a PCI-VME interface is also used for the high-speed data acquisition system. A Linux workstation with PENTIUM 4 CPU is the database server for the in-house developed database system. In this paper, the current status of SSTF DAC system, the database management system and recent test results are presented.

# 2 SSTF DAC SYSTEM

## 2.1 System Overview

Figure 1 shows a schematic diagram of SSTF DAC system. The basic configuration consists of operator interfaces, I/O controllers and Ethernet link. The operator interfaces are both UNIX workstations and Linux-based PCs. VME controller boards are based on Motorola 68040 CPU. VxWorks and RTEMS are used for the real-time OS of VME controllers. RT-Linux is the OS for the host computer of the high-speed data acquisition system, where SBS Model 620 is used for the PCI-VME interface and PENTEK 4275 is the main A/D converter. The data from 64 channels can be handled with 100 kHz sampling rate. Some independent devices such as strain gauge controllers, GPIB devices, vacuum furnaces and Helium liquefiers do not have an Ethernet interface. The data from such devices are collected to PCs and stored to an in-house developed database system using a NFS (network file system). Then, the data can be monitored using an in-house developed quasi-real-time monitoring system.
PENTEK 4270 DSP boards and PENTEK 4275 A/D converters with 4202 MIX baseboards are used for the quench detection and protection of superconducting magnet. Slow-varying parameters such as temperature, pressure, and Helium flow rate are monitored using VMIC VMIVME3122 A/D scanners. Various I/O modules such as VMIVME 4100 D/A Converter, PENTEK 1420 clock, and NIGPIB 1014 board are also installed in VME crates. Sets of PLC systems are used for the control of pneumatic valves, Helium refrigerator, vacuum furnace and etc. The National Instrument GPIB-ENET is also used for the control of GPIB devices such as a signal analyzer, a digital oscilloscope, voltage source, and etc.

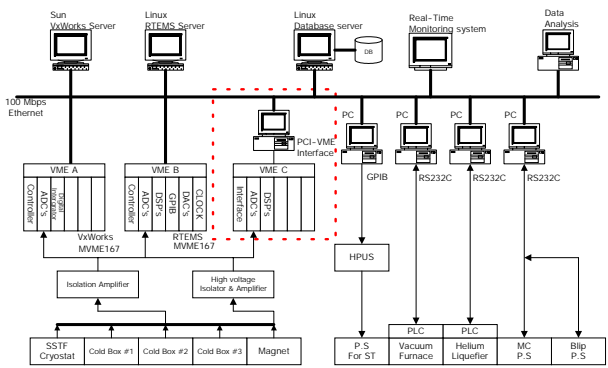

Fig.2 Current status of SSTF DAC system

Most of signals from various sensors are conditioned using isolation amplifiers. Isolation amplifiers have low pass filters and selectable gain controls by resistor change. A voltage tap signal could have a high voltage and high voltage isolation amplifiers are used to protect the data acquisition system with the maximum isolation voltage of 20 kV.

## 2.2 EPICS core Software

The main purpose of EPICS are to provide a fast, easy interface to data acquisition and control, and to provide an operator interface to all control system parameters.
At present, EPICS is being used as the data acquisition and control software only in VMIVME 3122 and VMIVME 4100 modules. The EPICS IOC database with respect to these I/O modules has been made using Capfast, which is a commercial schematic editor and compiled to be loaded into IOC via VxWorks startup script file. Each I/O controller provides a channel access server and both operator interfaces and I/O controllers are available as channel access clients. Using the client/server model and TCP/IP, channel access provides network-transparent access to the IOC database.
Most of parameters in the SSTF cryogenic system change slowly and the duration of the operation is the order of month. The EPICS is a convenient software tool for such an application and is used to control slow-varying parameters such as Helium flow rate, temperature, and pressure, etc.

## 2.3 Quench Detection and Protection

Quench is the phenomenon that a superconducting material is changing from a superconducting state to a normal conducting state. During the quench of a superconducting magnet, the magnet could be damaged both electrically and mechanically. In order to protect a magnet during quench, the data that is required for the quench detection is sampled with high speed during the operation of the magnet.
For example, the sampling rate of voltage tap signals is 100 kHz than the data is transmitted via local bus to the DSP board. According to the quench detection algorithm provided, DSP board analyzes the status of the superconducting magnet. In case of quench, DSP board generates a trigger signal for the quench protection system and the magnetic energy stored in the magnet is dissipated to the energy dump circuit.

## 3 DATABASE MANAGEMENT SYSTEM

The Linux OS with PENTIUM 4 CPU with 800 MHz Rambus DRAM is adopted as a database server because of its faster memory access time. The data from various sensors, GPIB instruments, vacuum gauge and power supply is stored with a given database format. The total number of data points is estimated to reach 5000 points in full-scale experiments.

The data has its own device name and data name. The data is stored in a separate file with binary format. Each file has three columns of double type data, which generally represent time index, raw measured values, and calibrated values, respectively. A certain data such as the voltage tap signal has the same raw values with calibrated values. The data from a device such as signal analyzer have the different data style and (time index, frequency, amplitude) and (time index, frequency, phase shift) will be stored in two separate file. Eventually, all the data during experiment can be easily accessed without modifying the application software for data retrieval and archival. Users merely have to know the data name and recorded date/time to retrieve the favorite data from the database. Also, users can download their data both in text and binary file format using another software. The text format is useful to perform analysis with the combination of graph applications running on PC platform. Besides, in order for users to monitor their equipment data more easily, two GUI programs are developed through x-window libraries. The application program, XY_MON, is a real-time data monitoring software using EPICS CA library and another is XYP which shows a graph after obtaining data through the database server access. XYP can also used for the quasi-real-time monitoring of the data through the continuous access of the database.

## 4 TESTS AND RESULT

Figure 3 shows the SSTF thermal shield cool-down history. XYP and XY_MON also generate the graphics output in postscript format, which is shown in Figure 3. The raw voltage data from the thermal shield of SSTF LVC is converted to the physical temperature data by built-in calibration routines in the database software, which also accesses the calibration data of various sensors.

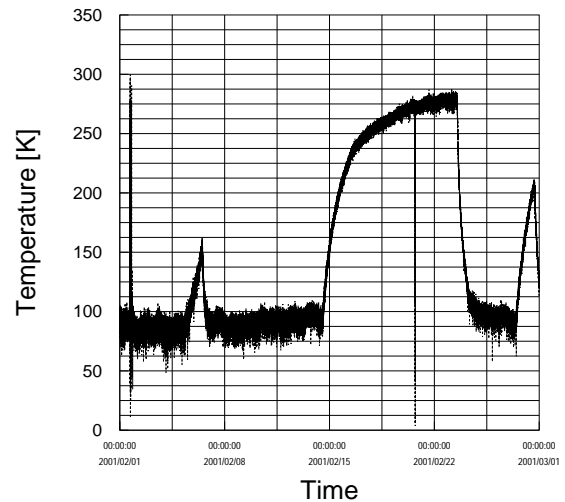

Fig.3 Thermal Shield Temperature of LVC

## 5 CONCLUSIONS

The database system with EPICS, which plays an essential role in the control system of SSTF, proved its reliability and scalability. For the full test of superconducting magnets under various conditions, the quench detection and protection system using DSP is still under development for more reliable quench detection algorithm. The data from the high-speed data logging system, which also include the data for the quench detection, is also stored to the database server with the same file format through DMA access between PCI adapter and VME adapter. RT Linux is currently adopted as basic OS for a real-time tasking in this system.